# Imaging stacking-dependent surface plasmon polaritons in trilayer graphene

Yilong Luan[1,2], Jun Qian[1,3], Minsung Kim[1,2], Kai-Ming Ho[1,2], Yi Shi[3], Yun Li[3], Cai-Zhuang Wang[1,2], Michael C. Tringides[1,2], Zhe Fei[1,2]*

[1]Department of Physics and Astronomy, Iowa State University, Ames, Iowa 50011, USA
[2]Ames Laboratory, U. S. Department of Energy, Iowa State University, Ames, Iowa 50011, USA
[3]National Laboratory of Solid-State Microstructures, School of Electronic Science and Engineering, Collaborative Innovation Center of Advanced Microstructures, Nanjing University, Nanjing 210093, People's Republic of China.

*Corresponding author: (Z.F.) zfei@iastate.edu

**Abstract:**

We report a nano-infrared (IR) imaging study of trilayer graphene (TLG) with both ABA (Bernal) and ABC (rhombohedral) stacking orders using the scattering-type scanning near-field optical microscope (s-SNOM). With s-SNOM operating in the mid-IR region, we mapped in real space the surface plasmon polaritons (SPPs) of ABA-TLG and ABC-TLG, which are tunable with electrical gating. Through quantitative modeling of the plasmonic imaging data, we found that the plasmon wavelength of ABA-TLG is significantly larger than that of ABC-TLG, resulting in a sizable impedance mismatch and hence a strong plasmon reflection at the ABA/ABC lateral junction. Further analysis indicates that the different plasmonic responses of the two types of TLG are directly linked to their electronic structures and carrier properties. Our work uncovers the physics behind the stacking-dependent plasmonic responses of TLG and sheds light on future applications of TLG and the ABA/ABC junctions in IR plasmonics and planar nano-optics.

**Main text:**

In recent years, graphene plasmonics has become an active research field due to the discovery of surface plasmon polaritons (SPPs) in single-layer graphene (SLG) with many superior properties such as high confinement, gate tunability, and a broad spectral range from terahertz (THz) to infrared (IR) [1-18]. Moreover, graphene could serve as a basic building block for the construction of a family of new plasmonic materials by van der Waals stacking. For example, it was found that AB-stacking bilayer graphene (BLG) supports gate-tunable IR SPPs [19], which could couple strongly with the intrinsic phonons of BLG [20] and exhibit peculiar reflection properties at the AB/BA domain walls [21]. Plasmonic studies have also been performed in twisted BLG, where SPPs have demonstrated sensitive dependence on the twist angle between the two graphene layers [22]. So far, experimental plasmonic studies have been focused mainly on SLG and BLG. The plasmonic responses of multi-layer graphene (MLG) with thicknesses beyond two layers are not fully explored.

One of the most popular MLG is trilayer graphene (TLG), which is formed by stacking three graphene layers together. TLG exfoliated from natural graphite has two common stacking orders: ABA (Bernal) stacking and ABC (rhombohedral) stacking [see insets of Fig. 1(b) and 1(c)]. The two stacking orders of TLG result in their dramatically different electronic structures. In Fig. 1(b) and 1(c), we plot the band structures of both ABA-stacked TLG (ABA-TLG) and ABC-stacked TLG (ABC-TLG) calculated with the tight-binding method [23,24]. Details about the calculations are given in the Supplemental Material [25]. The band structure of ABA-TLG [Fig.

1(b)] close to the charge neutrality point consists of a set of Dirac bands (labeled as 'v1') like those of single-layer graphene and a set of parabolic bands (labeled as 'v2') like those of bilayer graphene. Therefore, the low-energy carriers in ABA-TLG are a mixture of massless and massive carriers. The ABC-TLG [Fig. 1(c)], on the other hand, has a set of parabolic bands (labeled as 'v3') at low energies, so the carriers of ABC-TLG are massive. TLG has been studied extensively by transport, Raman, and far-field optical spectroscopies, where distinct electronic and phononic responses have been observed in the two types of stacking orders [26-34]. Recently, near-field imaging and spectroscopy have also been used to map in real space the stacking structures, domain walls, and hot-electron phonons of TLG with high spatial resolution [35-40].

In this work, we performed a systematic nano-IR imaging study of TLG to explore the stacking-dependent plasmonic responses. To image SPPs in TLG, we employed the scattering-type scanning near-field optical microscope (s-SNOM) operating in the mid-IR region. The s-SNOM was built based on an atomic force microscope (AFM), so we can obtain simultaneously the nano-IR and AFM topography images. For IR excitations, we utilized a continuous-wave (CW) $CO_2$ laser with a wavelength set to be 11.2 μm, corresponding to a photon energy of 0.11 eV. Upon laser illuminations, the metalized tip of the s-SNOM can efficiently launch and detect SPPs [14,15]. The samples studied here were obtained by mechanically exfoliating bulk graphite crystals onto the standard $SiO_2$/Si substrates. More detailed information about the experimental setup is given in the Supplemental Material [25]. In Fig. 1(d), we show an optical photo of a typical sample region that includes both ABA-TLG and ABC-TLG. We determined the thickness and the stacking orders of these samples by a combination of AFM and nano-IR imaging/spectroscopy via s-SNOM. With AFM, we measured the thickness of the sample to be about 1.0-1.2 nm uniformly throughout the sample [see Fig. 1(e) and inset]. With nano-IR imaging [Fig. 1(f)], we can visualize the two stacking orders: ABA-TLG has a higher IR signal compared to that of ABC-TLG as discussed in detail in previous literature [36,37]. Here and below, the signal shown in the s-SNOM images corresponds to near-field scattering amplitude ($s$) [14,15]. With nano-IR spectroscopy, we were able to measure the IR phonon resonance [40] that can also be used to distinguish the two stacking orders: ABC-TLG has a strong IR phonon resonance while ABA-TLG's IR phonon is too weak to be seen (see Fig. S1 in the Supplemental Material [25]).

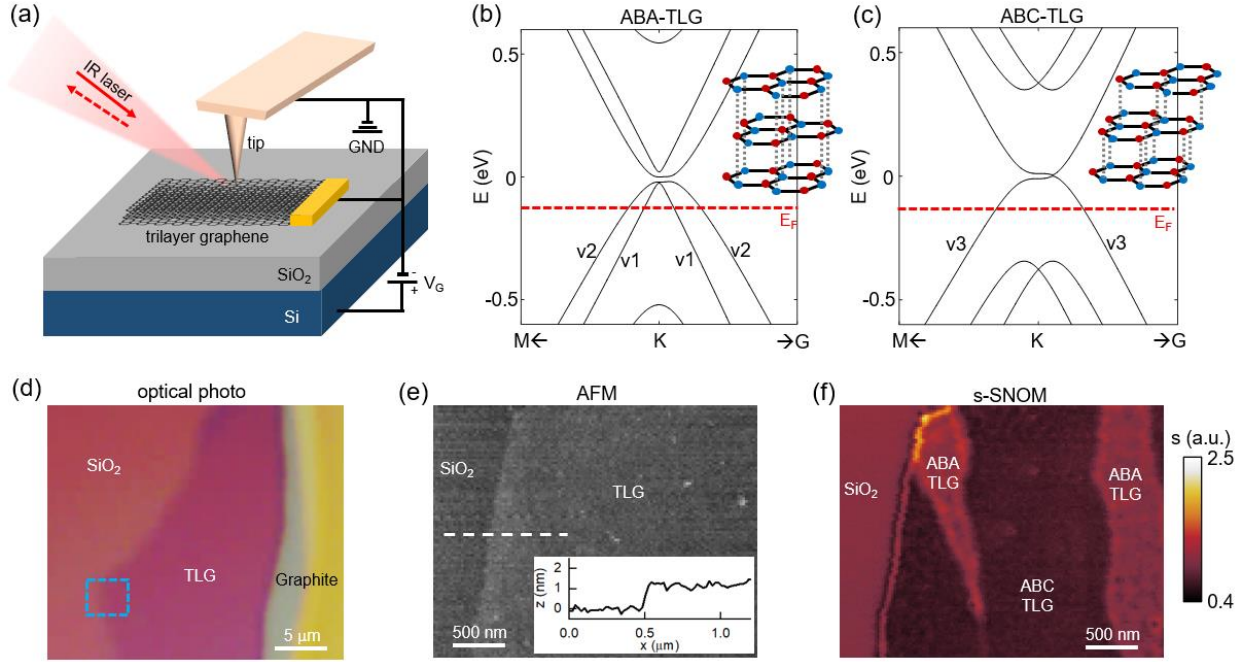

FIG. 1. (a) Illustration of the nano-IR imaging experiment of TLG with s-SNOM. (b),(c) The band structures and lattice structures (insets) of ABA-TLG and ABC-TLG, respectively. Here 'v1', 'v2', 'v3' mark the three sets of bands that are responsible for the plasmonic responses of the two types of TLG.  d, Optical photo of a TLG sample on a SiO$_2$/Si substrate. (e),(f) The AFM and nano-IR imaging data over the sample region marked in panel (d) (blue rectangle). The inset of panel (e) plots the AFM topography profile taken along the white dashed line. The nano-IR amplitude signal in (f) is normalized to that of the SiO$_2$/Si substrate.

From Fig. 1(f), we can see bright fringes close to the edges of the TLG sample. To visualize more clearly these fringes, we plot in Fig. 2 zoomed-in nano-IR images close to the sample edge taken at various back-gate voltages. Here, the voltages correspond to $V_g$ - $V_{CN}$, where $V_g$ is the applied gate voltage and $V_{CN}$ corresponds to the charge neutrality point ($V_g$ - $V_{CN}$ < 0 corresponds to the hole doping). As shown in Fig. 2, we can see bright fringes close to the edges of both ABA-TLG and ABC-TLG, which are generated due to the interference between tip-launched and edge-reflected SPPs [14,15,19,21]. These plasmonic fringes are stronger at higher doping (i.e., larger |$V_g$ - $V_{CN}$|) and almost completely disappear close to the charge neutrality point (i.e., $V_g$ - $V_{CN}$ = 0 V). Such gate dependence is a signature behavior of SPPs in graphene or other two-dimensional (2D) materials. Moreover, we found the fringes are different in the two types of TLG: plasmonic fringes are stronger and brighter in ABA-TLG than those of ABC-TLG.

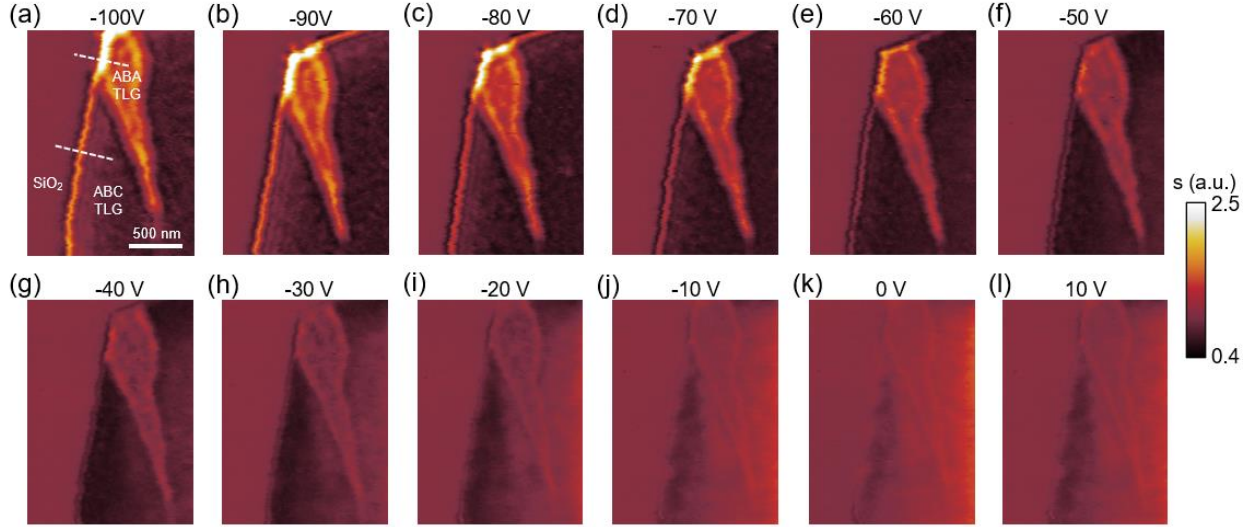

FIG. 2 The nano-IR imaging data of ABA-TLG and ABC-TLG at various gate voltages. The voltages marked here correspond to $V_g - V_{CN}$. The dashed lines in panel (a) mark the directions for extracting the line profiles shown in Fig. 3. The nano-IR amplitude signals plotted here are normalized to that of the SiO$_2$/Si substrate.

For quantitative analysis, we extract the line profiles perpendicular to the edges of ABA-TLG and ABC-TLG along the white dashed lines in Fig. 2(a). A selected set of line profiles (black curves) at different gate voltages are plotted in Fig. 3. Here the peaks in the profiles correspond to the bright fringes in the nano-IR images in Fig. 2. To extract the key plasmonic parameters of TLG, we fitted the line profiles with a quantitative model that was introduced in detail in previous works [14,19,21]. The key modeling parameter of the sample is the plasmonic wavevector $q_p$, based on which we can obtain the plasmon wavelength $\lambda_p \equiv 2\pi/\mathrm{Re}(q_p)$ and plasmon damping rate $\gamma_p \equiv \mathrm{Im}(q_p)/\mathrm{Re}(q_p)$. The modeling profiles are plotted as red dashed profiles in Fig. 3, which show good agreement with the experimental profiles. From the fitting, we determined $\lambda_p$ and $\gamma_p$ of both ABA-TLG and ABC-TLG, which are plotted in Fig. 4(a) and Fig. S2(a), respectively. We discuss mainly $\lambda_p$ of TLG in the main text (see Supplemental Material [24] for the discussions of $\gamma_p$). As shown in Fig. 4(a), $\lambda_p$ increases systematically with $|V_g - V_{CN}|$ for both ABA-TLG and ABC-TLG. This is expected since higher doping leads to higher conductivity and hence a larger $\lambda_p$. Moreover, we found $\lambda_p$ of ABA-TLG is much larger than that of ABA-TLG, which is the key origin for the distinct plasmonic and nano-IR responses of the two types of TLG [see Fig. 1(f) and Fig. 2]

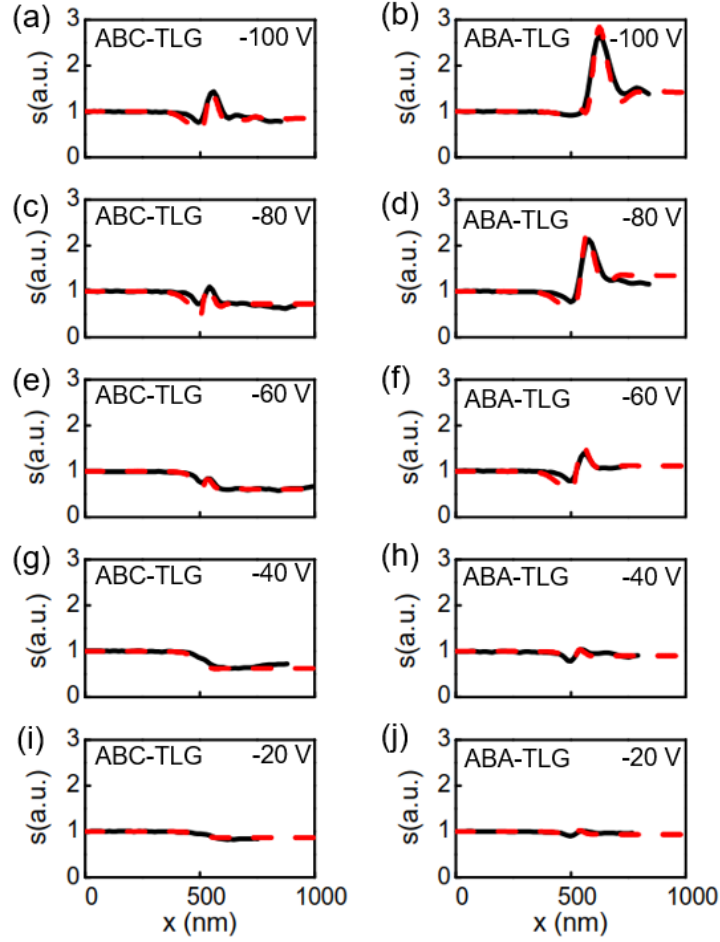

FIG. 3. Plasmon fringe profiles of ABC-TLG and ABA-TLG at various gate voltages ($V_g$-$V_{CN}$). The black curves are experimental line profiles taken perpendicular to sample edges (along white dashed lines) of the nano-IR images in Fig. 2. The red curves are the fitting profiles using a quantitative s-SNOM model [14,19,21]. The nano-IR amplitude signals in the profiles are normalized to that of the SiO$_2$/Si substrate.

The plasmon wavelength $\lambda_p$ is directly associated with the optical conductivity $\sigma(\omega)$ of the sample. Under the long-wavelength and 2D approximation, the plasmon wavevector ($q_p$) can be written as [14]

$$q_p \approx i\kappa\omega/2\pi\sigma(\omega). \quad (1)$$

Here $\kappa \approx (\varepsilon_S + 1)/2$ is the effective dielectric constant of the environment, and $\varepsilon_S$ is the permittivity of SiO$_2$. Under Drude approximation, $\sigma(\omega)$ can be written as $\sigma(\omega) = \sigma_0/(1-i\omega\tau)$, where $\sigma_0$ is the DC conductivity and $\tau$ is the carrier relaxation time. Therefore, the plasmon wavelength $\lambda_p \equiv 2\pi/\text{Re}(q_p)$ is roughly proportional to $\sigma_0$. Based on the Einstein relation, $\sigma_0$ has the following form:

$$\sigma_0 \approx \sum_{\text{all bands}} \frac{1}{2} e^2 N(E_F) v_F(E_F)^2 \tau. \quad (2)$$

In this equation, $N(E_F)$ is the carrier density at the Fermi level and $v_F$ is the Fermi velocity. The summation is over all the bands responsible for the conductivity [see Fig. 1(b) and 1(c)]. Based on

Eq. 1 and Eq. 2, we know that the plasmon wavelength $\lambda_p$ is roughly proportional to the DOS and $v_F^2$ at the Fermi level.

To determine DOS and $v_F^2$ at the Fermi level, we first calculated the Fermi energy $E_F$ at various gate voltages. As shown in Fig. 4(b), the $E_F$ of ABA-TLG is close to and slightly higher than that of ABC-TLG throughout the voltage region. We then calculate the total DOS and $v_F^2$ at the Fermi level for both types of TLG, which are plotted in Fig. 4(c) and 4(d), respectively. From Fig. 4(c), one can see that the total DOS of ABA-TLG is only slightly larger (≤ 20%) than that of ABC-TLG at high doping regime ($|V_g - V_{CN}| > 40$ V), but smaller at lower doping. A more dramatic difference can be seen in Fig. 4(d), where $v_F^2$ of the Dirac-like bands of ABA-TLG [band 'v1', see Fig. 1(b)] is significantly larger than that of the parabolic bands [band 'v2' of ABA-TLG and 'v3' and ABC-TLG, see Fig. 1(b) and 1(c)]. At high doping regime, $v_F^2$ of the Dirac bands is about 2 times that of the parabolic bands. The ratio increases to over 10 times at lower doping as the parabolic bands become more and more flattened. Based on Fig. 1(b) and 1(c), we believe that the big deviation of $\lambda_p$ in ABA-TLG and ABC-TLG showed in our data is mainly due to the difference in Fermi velocity. In other words, the Dirac carriers of ABA-TLG with higher $v_F$ are mainly responsible for the higher conductivity and larger $\lambda_p$ of ABA-TLG.

By calculating DOS and $v_F^2$ for every single band involved, we were able to compute the $\lambda_p$ of ABA-TLG and ABC-TLG using Eqs. 1 and 2 (Einstein relation), which are plotted in Fig. 4(a) as dashed curves. Calculations were also performed using the optical conductivity computed with the Kubo formula [solid curves in Fig. 4(a)] (see Supplemental Material [25] for details about the calculation). The results of the two theoretical methods agree with each other. From Fig. 4(a), one can see that both the doping and stacking dependence of the theoretical curves are qualitatively consistent with the experimental data points. We also notice that the calculated $\lambda_p$ of ABA-TLG is smaller (≤ 30%) than the experimental data points at the high doping regime. The causes of the deviation are not fully understood. It might be related to the tight-binding parameters (see Supplementary Material [25]) that we adopted from previous literature [32,40], which might not be ideal for our TLG samples.

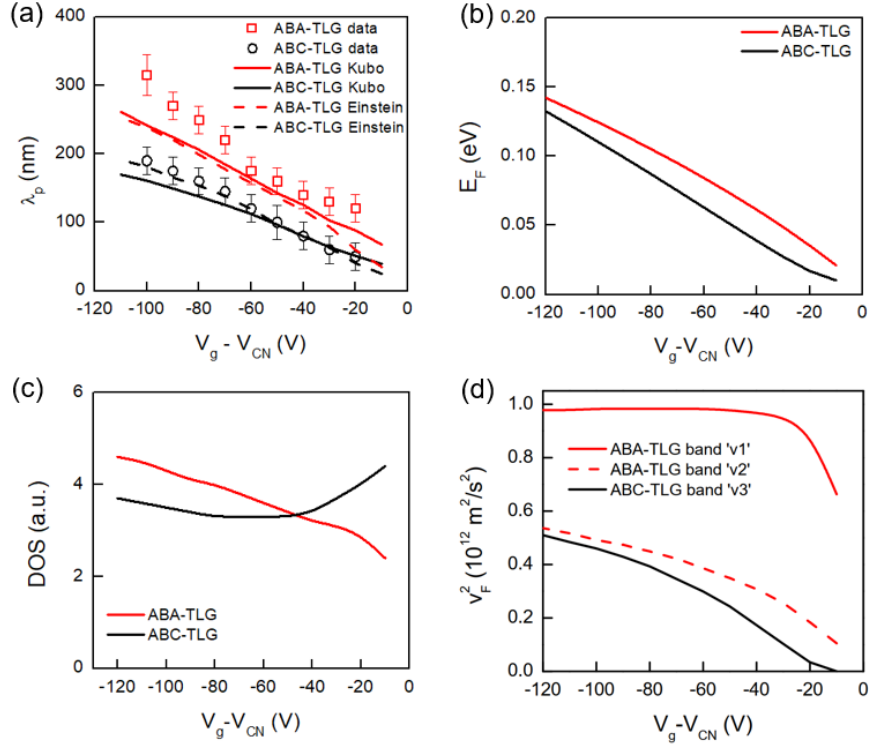

FIG. 4. (a) Experimental (data points) and theoretical (curves) plasmon wavelength $\lambda_p$ of ABA-TLG and ABC-TLG at various gate voltages ($V_g$-$V_{CN}$). (b) The calculated $E_F$ at various gate voltages for both ABA-TLG and ABC-TLG. (c) The calculated total DOS at various gate voltages for both ABA-TLG and ABC-TLG. (d) The calculated $v_F^2$ at various gate voltages for the Dirac-like bands 'v1' and parabolic-like bands 'v2' of ABA-TLG [see Fig. 1(b)] and for the parabolic bands 'v3' of ABC-TLG [see Fig. 1(c)].

Finally, we wish to discuss the plasmonic responses at planar junctions of ABA-TLG and ABC-TLG (termed as 'domain-wall solitons' in Ref. [35,44]. As shown in Figs. 1 and 2, bright plasmon fringes can also be seen at the ABA/ABC junctions indicating that they are efficient plasmon reflectors. This is primarily due to the big difference in $\lambda_p$ of the two types of TLG, leading to a sizable impedance mismatch of SPPs. Potential applications of these junctions include plasmon resonators where SPPs can be strongly confined and resonantly enhanced. One simple example is demonstrated in Fig. 5(a)-(c), where a special plasmon resonator formed by two ABA/ABC junctions is shown. Due to the plasmon reflection at the two junctions, one can see the plasmon fringe pattern evolves systematically with gate voltages and the width of ABA-TLG. At $V_g$-$V_{CN}$ = -100 V [see Fig. 5(a) and the fringe profiles in Fig. 5(d)], the number of plasmon fringes or peaks drops as the two junctions get closer to each other. Resonant enhancement is observed when the width of ABA-TLG drops to about 300 nm, where one strong plasmon fringe/peak is observed [see the black profile in Fig. 5(d)]. Similar plasmonic patterns have been studied previously in patterned graphene nanoribbons with the graphene edges as the plasmonic reflectors [14,15]. These graphene edges potentially have a lot of dangling bonds, so they are sensitive to chemical dopants and air molecules at ambient conditions. The ABA/ABC junctions of TLG, on the other hand, preserve intact crystal structures [36,45], so they are more robust and are less sensitive to the environment. Moreover, it has been demonstrated that these junctions can be physically engineered in a controllable way by a variety of physical or chemical methods [36,37,39,42-45]. Therefore, plasmonic resonators based on ABA/ABC planar junctions are in principle reconfigurable, which is not possible for those fabricated with lithography patterning and etching.

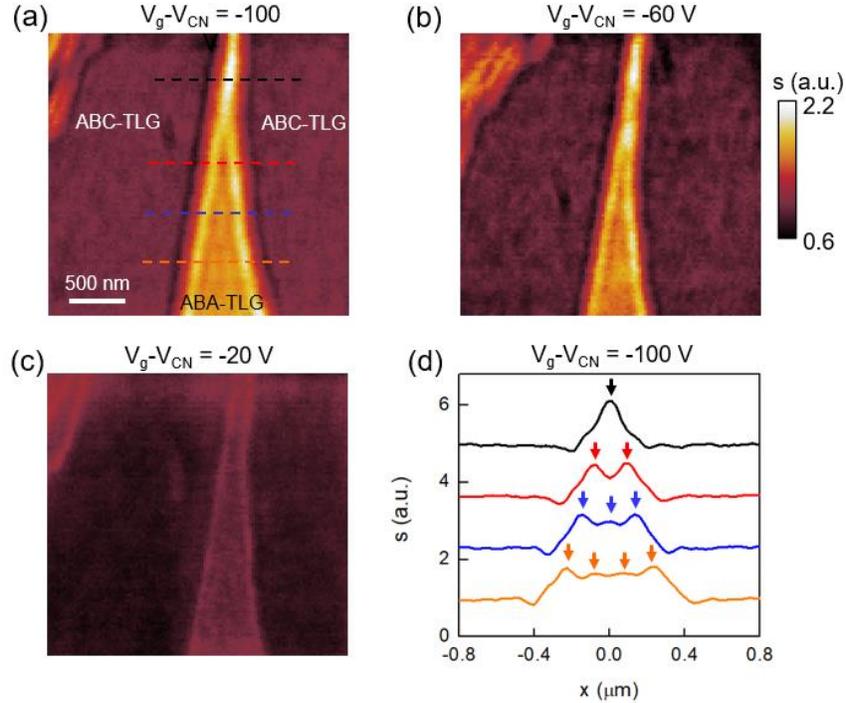

FIG 5. (a)-(c) Nano-IR imaging data of a sample area containing an ABC/ABA/ABC TLG junction at various gate voltages. (d) Line profiles extracted along the dashed lines in panel (a). The arrows mark the resonant peaks in the profiles.

In summary, we have performed a systematic nano-IR imaging study of SPPs in TLG with both ABA and ABC stacking orders. Through quantitative modeling of the interference fringe patterns due to SPPs, we found that the plasmon wavelength of ABA-TLG is significantly higher than that of ABC-TLG, which is mainly due to the larger Fermi velocity of Dirac carriers in ABA-TLG. Furthermore, we found that the planar junctions of ABA-TLG and ABC-TLG are efficient plasmonic reflectors. Plasmonic resonators formed by these junctions enable nanoscale localization and hence resonant enhancement of SPPs. Our work paves the way for future applications of TLG and ABA/ABC planar junctions in reconfigurable IR plasmonics and nano-optics.

**Acknowledgment**

This work is supported by Ames Laboratory. Ames Laboratory is operated for the U.S. Department of Energy by Iowa State University under Contract No. DE-AC02-07CH11358.